# Reductionism, emergence, and levels of abstractions
Russ Abbott
Computer Science, California State University, Los Angeles

Can there be independent higher level laws of nature if everything is reducible to the fundamental laws of physics? The computer science notion of *level of abstraction* explains why there can — illustrating how computational thinking can solve one of philosophy's most vexing problems.

More than six decades ago, Erwin Schrödinger [1] pondered the nature of life.

> [L]iving matter, while not eluding the 'laws of physics' … is likely to involve 'other laws,' [which] will form just as integral a part of [its] science.

Contrast Schrödinger's statement with this extract [2] from Albert Einstein.

> *The supreme test of the physicist is to arrive at those universal laws of nature from which the cosmos can be built up by pure deduction.*

Einstein represents strict reductionism: physics explains everything. Schrödinger says that there is more to nature than the laws of physics. But if biology is not just physics what else is there?

## *Unsolvability is a property of a level of abstraction*

The Game of Life is a cellular automaton in which cells are either alive (on) or dead (off). At each time step:
- any cell with exactly three live neighbors will stay alive or become alive;
- any live cell with exactly two live neighbors will stay alive;
- all other cells die.

The Game of Life rules are analogous to the fundamental laws of physics. They determine everything that happens on a Game of Life grid. Nevertheless there are higher level laws that are not derivable from them.

Certain Game of Life configurations create patterns. The most famous is the glider, a pattern of on and off cells that moves diagonally across the grid. It is possible to implement an arbitrary Turing machine by arranging Game of Life patterns. Computability theory applies to such Turing machines. Thus while not eluding the Game of Life rules, new laws (computability theory) that are independent of the Game of Life rules apply at the Turing machine level of abstraction — just as Schrödinger said.

Furthermore, conclusions about Turing machines apply to the Game of Life itself. Because the halting problem is unsolvable, it is unsolvable whether an arbitrary Game of Life configuration will reach a stable state.

Not only are there independent higher level laws, those laws have implications for the fundamental elements of the Game of Life. I call this *downward entailment,* a scientifically acceptable alternative to downward causation.

## *Evolution is a property of a level of abstraction*

Evolution by natural selection depends on (a) the possibility of heritable variation and (b) the effect of an entity's environment on the entity's ability to survive and reproduce. The more successful an entity is at surviving and reproducing, the more likely the features that define its relationship to its environment are to be passed on to its offspring. Variations that enable their possessors to survive and reproduce more effectively will propagate. Since the environment selects the features to be preserved, this is evolution by environmental (i.e., natural) selection.

Evolution is not a reductionist theory. It neither depends on nor is derived from lower level laws. Although reproduction and feature transmission are implemented by DNA, Darwin and Wallace didn't know about DNA. They didn't have to. Evolution occurs in any level of abstraction that includes heritable variation and environmentally influenced survival and reproduction.

## *The reductionist blind spot*

Game of Life patterns and evolution are both epiphenomenal — they have no causal power. Consider gliders. They don't do anything. It is only the Game of Life rules, not gliders, that make cells go on and off. When describing the Game of Life, one can always reduce away macro-level patterns like gliders and replace them with the underlying micro phenomena. The same is true for evolution. One can always describe how a population came into being in terms of DNA and other lower level mechanisms. It is always the elementary mechanisms that turn the causal crank. So why not reduce away epiphenomenal levels of abstraction?

Reducing away a level of abstraction results in a reductionist blind spot. No set of equations over the domain of Game of Life grid cells can describe the computation performed by a Game of Life Turing machine — unless the equations themselves model a Turing machine. If one takes reduction seriously, all one may talk about are elementary objects, e.g., grid cells. At that level there are no Turing machines — and no computability theory. The laws that characterize regularities at higher levels of abstraction become impossible to express when the abstractions are reduced away.

Furthermore, levels of abstraction are objectively real. They have observably reduced entropy. Game of Life patterns and biological organisms can be described much more compactly than by enumerating cell states and elementary particles.

In addition, entities at higher levels of abstraction have mass properties that differ from those of their components. Static entities (atoms, molecules, solar systems — entitles at an energy equilibrium) have less mass than their components taken separately. Dynamic entities (biological organisms, social organizations, hurricanes — far-from-equilibrium entities that must extract energy from their environment to persist) have more mass than their components.

The only forces in nature are the fundamental forces of physics; all higher level force-like interactions are epiphenomenal. But there is more to nature than forces. The answer to "What else is there?" is: levels of abstraction. The goal of science is to understand and explain nature at all levels. Reducing away objectively real and explanatorily powerful levels of abstractions is bad science.

Two definitions sum this up.

- **Emergence**: the implementation — either statically (at equilibrium) or dynamically (far from equilibrium) — of a level of abstraction. Formation or dissolution of a level of abstraction often manifests as a phase transition.

- **Generalized evolution**: the principle that extant levels of abstraction (naturally occurring or man-made) are those whose implementations have materialized and whose environments support their persistence.

It shouldn't be surprising that levels of abstraction obey new laws. A level of abstraction is implemented by constraining some underlying system. A constrained system obeys laws that do not apply to the unconstrained system. But these new laws are creative additions to those that apply at lower levels. They are creative in the same sense that software is creative. We write software to create new worlds — that obey new laws. Nature too is a programmer — a blind programmer.

This discussion illustrates how computational thinking can help solve a fundamental problem in the philosophy of science. Computational thinking extracts lessons from software — which is grounded in reality in a way that purely abstract disciplines like mathematics are not. To be successful, software must execute. Science and software are both products of the mind whose ultimate test is in the world. For additional discussion see [3] and [4].

**References**

[1] Schrödinger, Erwin, *What is Life?*, Cambridge University Press, 1944.

[2] quoted in Gross, David, "Einstein and the search for unification," Current Science, 89/12, 25 December 2005, pp. 2035 – 2040.

[3] Abbott, Russ, "Emergence explained," *Complexity*, Sep/Oct, 2006, (12, 1) 13-26. Preprint: http://cs.calstatela.edu/wiki/images/9/95/Emergence_Explained-_Abstractions.pdf.

[4] Abbott, Russ, "If a tree casts a shadow is it telling the time?" to appear in the *Journal of Unconventional Computation*. Preprint: http://cs.calstatela.edu/wiki/images/6/66/If_a_tree_casts_a_shadow_is_it_telling_the_time.pdf.